# Evidence for 1D chiral edge states in a magnetic Weyl semimetal Co$_3$Sn$_2$S$_2$

Sean Howard[1], Lin Jiao[1], Zhenyu Wang[1], Noam Morali[2], Rajib Batabyal[2], Pranab Kumar-Nag[2], Nurit Avraham[2], Haim Beidenkopf[2*], Praveen Vir[3], Chandra Shekhar[3], Claudia Felser[3], Taylor Hughes[4], Vidya Madhavan[1*]

[1]Department of Physics and Frederick Seitz Materials Research Laboratory, University of Illinois Urbana-Champaign, Urbana, Illinois 61801, USA
[2]Condensed Matter Physics Department, Weizmann Institute of Science, 7610001 Rehovot, Israel
[3]Max-Planck-Institute for Chemical Physics of Solids, 01187 Dresden, Germany
[4]Department of Physics and Institute for Condensed Matter Theory, University of Illinois at Urbana-Champaign, Urbana, Illinois 61801, USA

**Abstract**

The physical realization of Chern insulators is of fundamental and practical interest, as they are predicted to host the quantum anomalous Hall effect (QAHE) and topologically protected chiral edge states which can carry dissipationless current. Current realizations of the QAHE state often require complex heterostructures and sub-Kelvin temperatures, making discoveries of intrinsic, high temperature QAH systems of significant interest. In this work we show that time-reversal symmetry breaking Weyl semimetals, being essentially stacks of Chern insulators with inter-layer coupling, may provide a new platform for the higher temperature realization of robust chiral edge states. We present combined scanning tunneling spectroscopy and theoretical investigations of the recently discovered magnetic Weyl semimetal, Co$_3$Sn$_2$S$_2$. Using modeling and numerical simulations we find that depending on the strength of the interlayer coupling, chiral edge states can be localized on partially exposed kagome planes on the surfaces of a Weyl semimetal. Correspondingly, our dI/dV maps on the kagome Co$_3$Sn terraces show topological states confined to the edges which display linear dispersion. This work provides a new paradigm for realizing chiral edge modes and provides a pathway for the realization of higher temperature QAHE in magnetic Weyl systems in the 2D limit.

The quantized Hall conductance of the quantum Hall effect is a striking example of the macroscopic consequences of quantum phenomena[1]. In the quantum Hall effect, large magnetic fields generate Landau levels in a 2D material. The Landau levels acquire a non-zero topological index, resulting in chiral edge currents that are a manifestation of the quantized Hall response. Haldane's conception of the Chern insulator[2], or quantum anomalous Hall insulator[3], takes this idea a step further. A Chern insulator is a 2D material that exhibits the quantum Hall effect in the absence of an external magnetic field. The distinctive features of Chern insulators are their quantized Hall conductance, and topologically protected chiral edge states[4], which travel in unidirectional channels (see Fig. 1a). The transport signatures of the QAHE were originally reported in a 2D magnetically doped topological thin film[5], and advancements in QAHE signatures in magnetically doped topological insulators[6–13] have recently been extended to intrinsic magnetic topological insulators[14] and twisted bilayer graphene[15]. While these results are breakthroughs in studying QAHE, future progress in studying chiral edge states is limited by the low Curie temperatures (< 30 K) of these material systems and the complex heterostructures often necessary to realize chiral edge states.

Interestingly, Chern insulators are also related to a variety of higher-dimensional topological systems, not least of which are the magnetic Weyl semimetals (WSMs). One can in fact model magnetic WSMs as layers of 2D Chern insulators that are coupled in the stacking direction [16,17]. Thus, an unexplored route to 2D Chern insulators is to identify a 3D layered magnetic Weyl semimetal fitting this description. Recent developments in candidate magnetic WSMs[18,19] now provide a promising alternative arena for the study of chiral edge states. In the spirit of the coupled-layer model presented by Balents and Burkov[16], we can show that stepped terraces on the surface of a WSM can harbor chiral edge states localized on the steps. First, we note that Weyl semimetals are an intermediate critical phase between a trivial insulator and a magnetic weak topological insulator, the latter of which is adiabatically connected to a decoupled stack of Chern insulators[20–22]. Importantly, the two gapped phases and the intermediate gapless phase can be reached by starting with decoupled layers of Chern insulators and increasing the strength of the inter-layer coupling. To illustrate, take a bilayer of Chern insulators both having the same non-zero Chern number. If the coupling between the two layers is increased, the system will eventually undergo a transition where the strong tunneling creates a trivial phase of the bilayer with vanishing Chern number. Interestingly, if we strip off a part of one of the layers as depicted in the schematic in Fig. 1b, the exposed single-layer region will revert to a non-trivial Chern insulator since it was the inter-layer coupling that drove it to be trivial. Thus, both the end of the single layer, and the single-step defect itself will harbor a chiral edge state (see Supplemental information 1 for more details). The existence of the edge state on the end of the single layer is obvious, but the edge state on the step defect appears only because the remaining bilayer region with strong tunneling is a trivial insulator. Thus, this process serves to expose a region of Chern insulator despite the full bilayer system being trivial. The concept of exposing topological sub-systems when the combined system is trivial was explored in two recent papers[23,24] in the context of a bulk topological proximity effect, and embedded topological insulators respectively.

While a single bilayer of Chern insulators is not sufficient to realize the intermediate WSM phase between the topological and trivial regimes, we extend this analysis to the case of many layers (as shown schematically in Fig. 1c) to model the properties of a WSM (Supplemental Information 1). We find that step-localized chiral edge states continue to exist in WSMs in a wide swath of the topological phase diagram that is parameterized by the Chern-insulator gap of a decoupled QAHE plane, and the inter-layer tunneling (see Supplemental Information 1). Indeed, we find that whenever the surface terrace exhibits the localized chiral modes on the steps, the system is in a WSM phase. Thus, while not every Weyl semimetal will harbor step-localized chiral modes, a large fraction do. One can therefore be optimistic that, given a magnetic WSM, there is a large probability that terraces will exhibit localized QAHE regions and manifest localized chiral modes.

To investigate these theoretical predictions, we study the magnetic Weyl semimetal $Co_3Sn_2S_2$. There is substantial prior evidence for the topological WSM nature of this material, including a large anomalous Hall effect[25,26], signatures of Fermi arc states in STM[27–29], and flat band diamagnetism caused by Berry curvature[30]. Importantly, the compound is predicted to host the QAHE in the 2D limit[31] and models of a single, magnetic $Co_3Sn$ kagome layer predict a non-zero Chern number[30]. Therefore, this material appears to fit the model of a WSM constructed from stacked and coupled Chern insulators. Additionally, the material's high Curie temperature of 170 K[32] makes $Co_3Sn_2S_2$ an ideal candidate to test the predictions of terraces within the coupled Chern insulator model. If edge states are observed, then it would provide strong experimental evidence that the 2D limit of this material may host the QAHE at elevated temperatures.

We use scanning tunneling microscopy and spectroscopy (STM/S) at 4 K to investigate the possibility of realizing chiral edge states in $Co_3Sn_2S_2$. $Co_3Sn_2S_2$ is a layered material consisting of a kagome $Co_3Sn$ plane in-between two hexagonal S layers, and all sandwiched in between two hexagonal Sn layers (Fig. 1d-f). The hexagonal lattice constant is a = 5.3 Å, while three stackings of Sn-S-$Co_3Sn$-S-Sn layers each translated by (1/3,1/3,1/3) construct a full unit cell, giving a lattice constant c = 13.2 Å[19]. The material is a half-metallic ferromagnet, with magnetic properties derived from the moments of the Co atoms aligned along the c-axis[18,32].

$Co_3Sn_2S_2$ bulk single crystals cleaved along the (001) direction most often expose two distinct surfaces[28–30] that are both hexagonal in nature. This suggests that the main cleavage plane is between the S-Sn/Sn-S layers revealing either the Sn or S layer. A third possible termination is the honeycomb like kagome $Co_3Sn$ plane which is rare. The two surfaces most commonly seen in STM can be distinguished topographically (one consisting of vacancies and the other adatoms) and also spectroscopically (as shown in Fig. 1g,h). On the surface with adatoms, we observe a sharp peak in the spectra near -12 meV associated with a diamagnetic flat band[30], along with another large peak near -300 meV. On the surface with vacancies, we see a depression in the density of states from -300 meV to 0 meV, with two broad peaks occurring at 50 meV and 200 meV respectively.

Previous STM studies have arrived at different conclusions on the chemical identification of these two common hexagonal surfaces[28–30]. To reconcile this issue, we utilize the symmetry of the local density of states signatures of defects to identify the termination layer. This method has been used successfully for chemical identification of surface lattices of other layered materials[33,34]. Density of state signatures of defects centered at positions that do not correspond to the positions of the top layer atoms can be attributed to defects in the layer below (DLB). Such signatures are present in both of the commonly observed hexagonal surfaces. However, DLBs existing on the surface with vacancies have a reduced symmetry (a triangle with one bright vertex) than the symmetry of DLBs on the surface with adatoms (a clover with equally bright vertices). The reduced symmetry of DLBs seen on the surface with vacancies is consistent with the layer below being the $Co_3Sn$ kagome plane, which has the same reduced symmetry compared to hexagonal layers. This allows for the identification of the surface with vacancies as being the

S plane, and the surface with the adatoms being the Sn plane (see Supplemental Information 2) which is consistent with previous a STM study incorporating DFT calculations[28].

To look for edge modes, we investigate large $Co_3Sn$ terraces terminated by a step edge (Fig 2a). The $Co_3Sn$ surface can be distinguished spectroscopically from the S and Sn planes. From Fig 2b, the spectra seen on the $Co_3Sn$ surface have two sharp peaks in the density of states at 0 meV and +60 meV which are not seen in the spectra on either the S and Sn surfaces and are consistent with spectra previously reported on the $Co_3Sn$ surface[28,29]. Point spectra taken on the edge of this surface reveals a broad accumulation density of states that is not present in spectra taken away from the edge. The broad accumulation of density of states at the edge over these energies is unique to the $Co_3Sn$ surface (see Supplemental Information 3). Taking spectra along the line profile shown in Fig 2a reveals that this enhanced density of states is localized to the edge, with an extent of approximately 1.5 nm transverse to the step edge (Fig 2c and Supplemental Information 4). The highly localized nature of these states suggests that they can be attributed to an edge mode existing on an exposed $Co_3Sn$ plane. Interestingly, despite the presence of impurities on the edge which act as scattering centers, we find no sign of quantization in the dI/dV spectra of the edge mode (see Supplemental Information 4). This however changes markedly when we investigate narrow terraces, as we show next.

To further explore the nature of these edge states, we study narrow terraces of the kagome $Co_3Sn$ planes. Incomplete cleavage between Sn and S layers occasionally exposes small terraces near step edges on the S surface (Fig. 3a and Supplemental Information 5,6). These terraces are approximately 130 pm below the S surface (Fig. 3b). As shown in the extended data (Supplemental Information 7), the closest plane to S is the $Co_3Sn$ plane at a height of 156 pm below the Sn surface with the next plane being 312 pm below. The thin terrace can thus be unambiguously identified as the $Co_3Sn$ plane. Taking a dI/dV map of one of these terraces reveals striking features, highlighted by quantum well like bound states at various energies shown in the DOS images in Fig. 3c-f. Note that these states are responsible for obscuring the atomic resolution on this terrace. The number of nodes increases with energy, indicating a positive dispersion. The bound state energies for each distinct quantum well like state can be identified from peaks in the spectra at locations where the density of states is maximum (see Fig. 3g). In the terrace of length 5.2 nm, shown in Fig. 3a, we observe a sequence of four quantum well like states ranging from $n$=1 to $n$=4, while in the other two terraces, with lengths of 5 nm and 6.1 nm as shown in Supplemental Information 3, we observe a sequence of four and five quantum well like states, respectively. The notation for the $n^{th}$ state is the same as in the classic quantum well, where $n$ indicates a bound state with $n$+1 nodes and $n$ maxima in the local density of states. Converting the linear energy dependence on the bound state wavelength into a dispersion velocity we find a value of approximately $5 \times 10^4$ m/s. Quantum well like 1D states have been observed previously with STM in systems such as Au/Cu adatom chains and semiconductor terraces[35–39], however all of these studies show a quadratic dependence of the bound state/sub-band energy on the number of nodes, as expected from conventional quantum well states originating from free-electron-like quadratic dispersion. Unlike these examples, the dispersion seen on all the terraces we observe is linear (as plotted in Fig. 3h).

To interpret the STM results using the layered WSM model (Supplemental Information 1) we first perform numerical simulations of a WSM having a surface terrace with a partially exposed Chern insulator plane as shown in Fig. 4a top panel. Spacer layers between Chern insulator layers, in this case the S and Sn layers, effectively change the coupling in the stacking direction. Identifying S/Sn as spacer layers is consistent with earlier work that shows that the band structure near the Fermi energy mainly arises from the Co-3d bands[26] as well as work showing that the Kagome monolayer alone is a Chern insulator[30]. Remarkably, we find that the chiral modes that we argued are present in the strongly coupled bilayer are recovered in the WSM regime, as seen in Fig. 4a bottom panel. These modes counter-propagate and are localized at edges of the partially exposed Chern insulator planes and decay exponentially along the surface and as a power law into the bulk.

After confirming the existence of the chiral terrace modes within the coupled Chern insulator model, we further simulate the DOS signatures of the two (counter-propagating) chiral edge states in a quantum well using a simple one-dimensional lattice model for linearly dispersing, counter-propagating modes in a potential well. We use a one-dimensional Hamiltonian of the form $H = \sum_n \left( \frac{i}{2} c_{n+1,\alpha}^\dagger c_{n,\beta} \sigma_{\alpha\beta}^z - \frac{i}{2} c_{n,\alpha}^\dagger c_{n+1,\beta} \sigma_{\alpha\beta}^z + m c_{n,\alpha}^\dagger c_{n,\beta} \sigma_{\alpha\beta}^x + V(n) c_{n,\alpha}^\dagger c_{n,\beta} \delta_{\alpha\beta} \right)$. Here the operator $c_{n,\alpha}^\dagger$ creates an electron on site $n$ in orbital α, the parameter $m$ represents the evanescent hybridization between the counter-propagating modes, $V(n)$ is a discretized finite square well potential, and all energy scales are in units of the tunneling strength which we have set to unity. See Fig. 4a,b and Supplemental Information 8 for a schematic illustration of the model setup. In the absence of the square well potential, this Hamiltonian yields an energy spectrum $E = \pm\sqrt{\sin(k_a)^2 + m^2}$ which, for vanishing hybridization ($m$=0) and long wavelengths ($k_a$<<1), recovers the linear dispersion $E = \pm|k|$ (see left most panel in Fig. 4c and Supplemental information 8). Turning on the hybridization term $m$ opens a gap, lifting the degeneracy at zero momentum as shown in the middle panel of Fig. 4c. The length of the terrace is modeled as a finite square well, $V(n)$, with a width that is a small fraction of the total system size. Adding a finite potential well with a non-zero mixing term creates eigenstates within the hybridization-induced energy gap, as illustrated in the right most panel in Fig 4c. The electron density of the eigenstates confined to the potential well resembles the bound states observed in our experiment (see Fig. 4d). Additionally, for a wide range of $m$, the energies of the confined states are linearly dependent on the confinement quantum number $n$ (see Fig. 4e and Supplemental Information 8). Also, if $m$ becomes too large, the dispersion relation begins to resemble a typical quadratic band and the bound state energies cross over to a quadratic dependence on the confinement quantum number. If $m$ is too small the counter-propagating modes will not effectively form a bound state, e.g., if the plateau is very wide so that the evanescent coupling between the opposed edge modes is small, the chiral edge mode will hit the potential wall and turn to continue its circulation around the plateau boundary instead of forming a coherent bound state. Physically, this indicates that when the terrace width is small enough to hybridize chiral edge states, we should expect quantum well like states to develop. Notably, the transverse extent of the edge state observed on the $Co_3Sn$ surface is of the same order (~1.5 nm) as the width of terraces containing quantum well like bound states (Fig. 2 and Supplemental Information 5). This suggests that the edge states are within the intermediate evanescent mixing regime necessary to observe linearly dispersing bound states.

We have found that our models can reproduce all the features seen in our experiment, providing a clear, self-consistent explanation for the existence of linearly dispersing quantum well like bound states, composed of hybridized chiral edge states on an exposed kagome $Co_3Sn$ terrace. While the chiral nature of these states cannot be directly probed with STM, it follows naturally from topological edge states within a material with broken time-reversal symmetry. It is worth noting that similar interference patterns of topological edge states have been observed by STM in Bi single crystals[40,41]. While Bi edge state interference patterns are due to intra-band scattering of topological edge states, our result originates from chiral edge states with the same quantum numbers that couple because of their close spatial proximity.

This observation of chiral edge modes within a bulk magnetic Weyl semimetal provides evidence for the physical realization of the model presented by Balents and Burkov[16]. The modification of this original model to terrace geometries, combined with our experimental observations, theorizes a new paradigm for studying chiral edge states in a wide range of magnetic Weyl semimetals via local probes without requiring thin film growth or heterostructures. Most importantly, a reductionist approach to this model suggests that a material fitting this description will, in the 2D limit, be a Chern insulator hosting the QAHE. $Co_3Sn_2S_2$'s high Curie temperature[32] which persists into the 2D limit[42], existing theoretical calculations in the 2D limit[31], and our observation of linearly dispersing chiral edge modes make the 2D limit of $Co_3Sn_2S_2$ a strong candidate for the observation of intrinsic QAHE at elevated temperatures.

## Data availability

The data in this work will be made available on reasonable request. Correspondence and requests for materials should be addressed to V.M.

**Figure 1 | Crystal Structure, Topography, and Spectroscopy of $Co_3Sn_2S_2$**

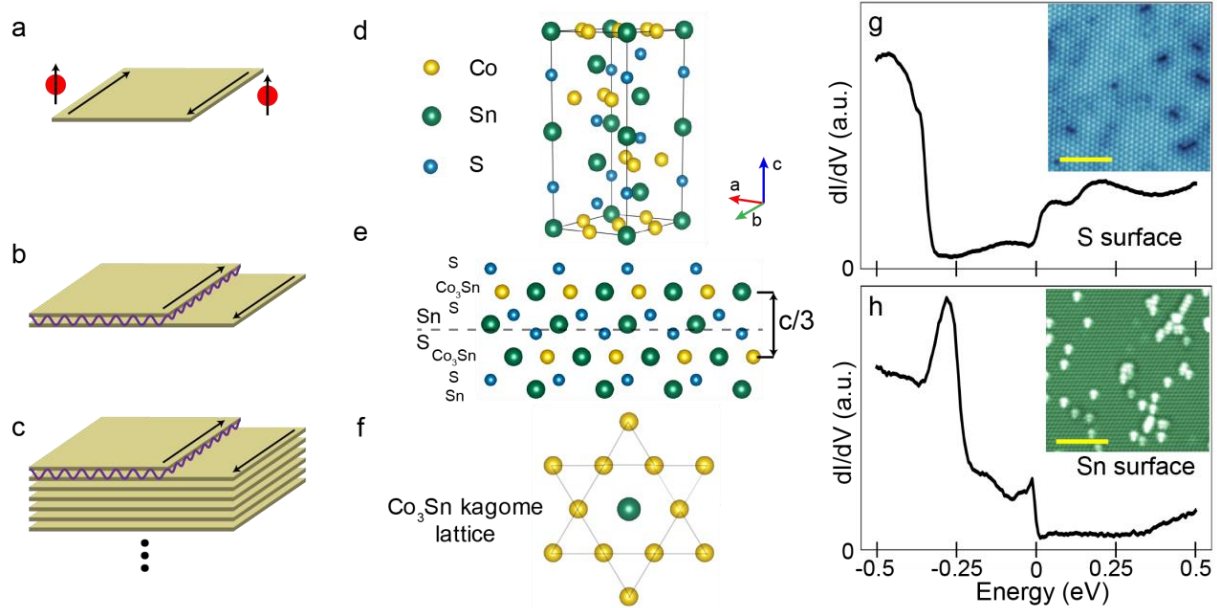

**a,** Single Chern insulator, with one chiral edge state. **b,** Chern insulator bilayer system with step geometry and inter-layer coupling (represented by the purple lines). If the coupling is strong enough to turn the bilayer system trivial, chiral edge states will appear at the edge of the exposed plane and the interface of the bilayer and single layer. **c,** Large stack of Chern insulators with intermediate inter-layer coupling to form a Weyl semimetal. For large regions of parameter space, the Weyl semimetal phase coincides with the conditions necessary for the counter-propagating chiral edge states seen in **b** (see Supplemental Information 1). **d,** Unit cell of $Co_3Sn_2S_2$. Sn is represented by large green sphere, Co by medium yellow sphere, and S by small blue sphere. **e,** Several layers of the crystal as viewed from the (100) plane. Cleavage occurs between the Sn and S layers, as indicated by a dashed line. **f,** Schematic of the kagome structure present in the $Co_3Sn$ plane with Co forming 6 triangles surrounding a central Sn atom. **g,** Spectra typical of S surface. Inset is 30 nm x 30 nm topography showing vacancies typical of S surface. **h,** Spectra typical of Sn surface. Inset is 30 nm x 30 nm topography showing adatoms typical of Sn surface. The yellow lines in the insets of g and h are 10nm scale bars.

**Figure 2 | Spectroscopic Data Showing Edge State on Co₃Sn plane**

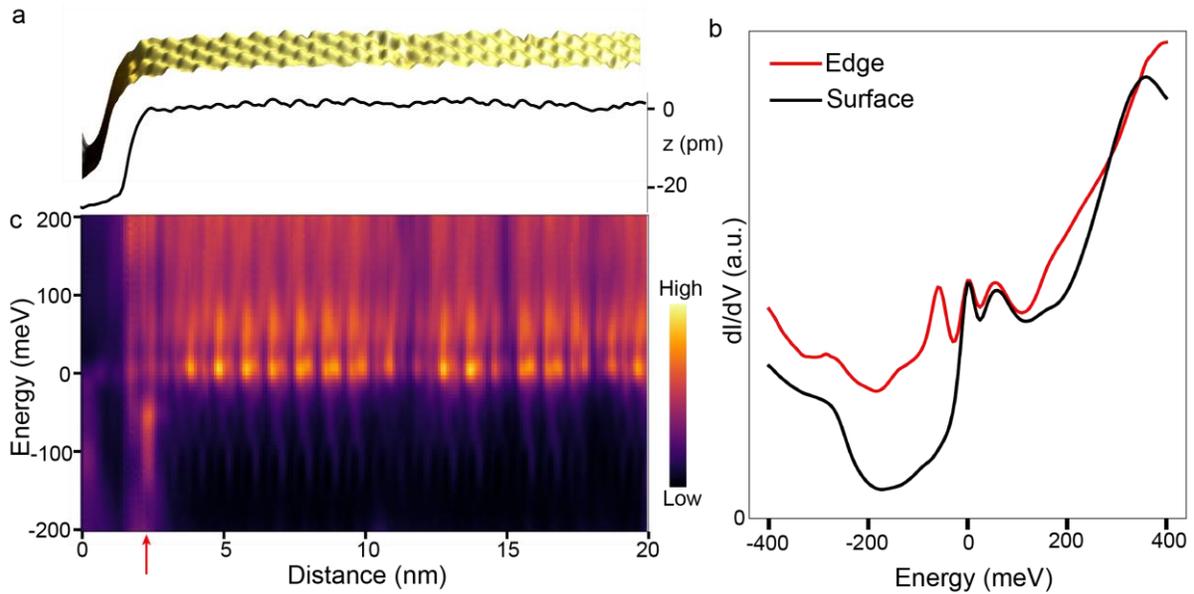

**a,** Three-dimensional topography and line profile of an exposed Co₃Sn surface. Only part of the step is shown here. The atomic scale features represent a cluster of three nearest neighbor Co-atoms. lThe full step and zoomed out image are shown in the Supplement Fig.4. **b**, Spectroscopic heatmap of the differential conductance along the line profile shown in **a**. An edge state is clearly seen as indicated by the red arrow. The state is confined to the edge with an extent of approximately 1.5 nm transverse to the step edge. **c,** Spectra on the Co₃Sn plane taken at the edge (red) and away from the edge on the surface (black). A noticeable peak in the density of states at the edge exists at -60 meV along with a broad increase in the density of states below the Fermi energy, indicative of the edge mode.

**Figure 3 | Observation of Linearly Dispersing Bound States on Co$_3$Sn Kagome Terrace**

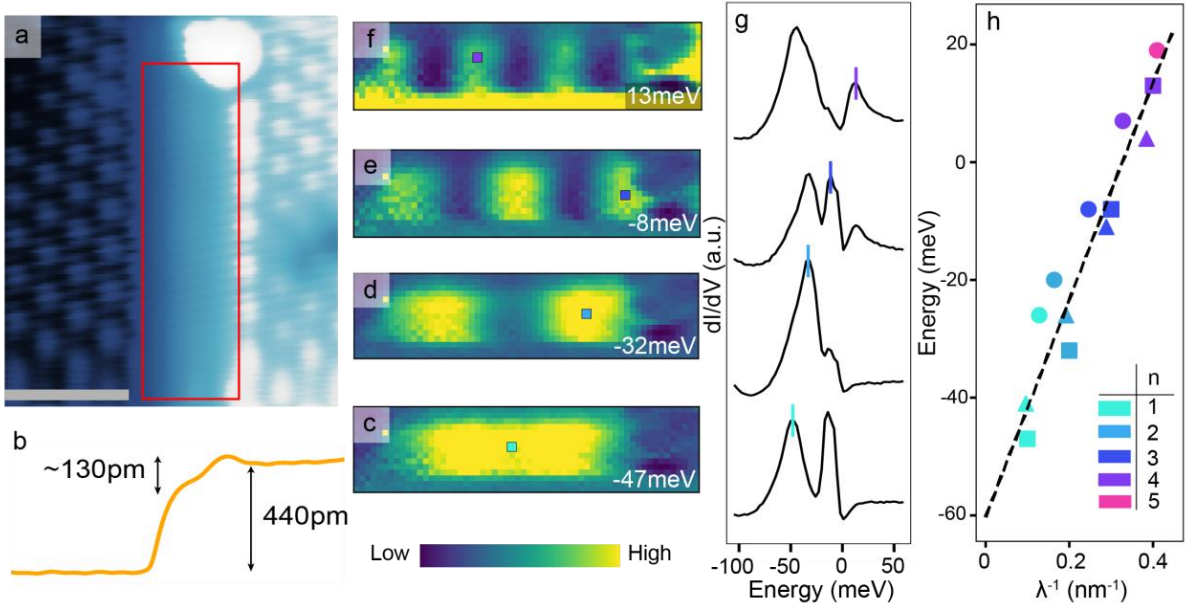

**a**, Topography of a step edge between two S surfaces containing a small terrace of the Co$_3$Sn plane, indicated by the red rectangle. Gray inset scale bar indicates 2nm. **b**, Line profile of the step edge. Total step height is consistent with one third of a $\vec{c}$, or approximately 440pm. The terrace is approximately 130pm below the top S surface. **c-f,** Density of state maps of the region enclosed by red rectangle in **a** at energies for each quantum well like state. **g**, Spectra at locations of high density of states for each quantum well like state indicated by small squares in **c-f** starting at n=1 at the bottom to n=4 at the top. Spectra are offset for clarity. A small vertical line indicates the peak associated with energy of state. **h,** Energy versus inverse wavelength relation for states found on three Co$_3$Sn terraces. The states shown in figure are represented by squares, while states from the two other terraces are circles and triangles. The linear dashed line is provided as a guide to the eye. The inverse wavelength is calculated assuming the distance between peaks is one wavelength.

**Figure 4 | Tight Binding Calculation of Linearly Dispersing States within a Potential Well**

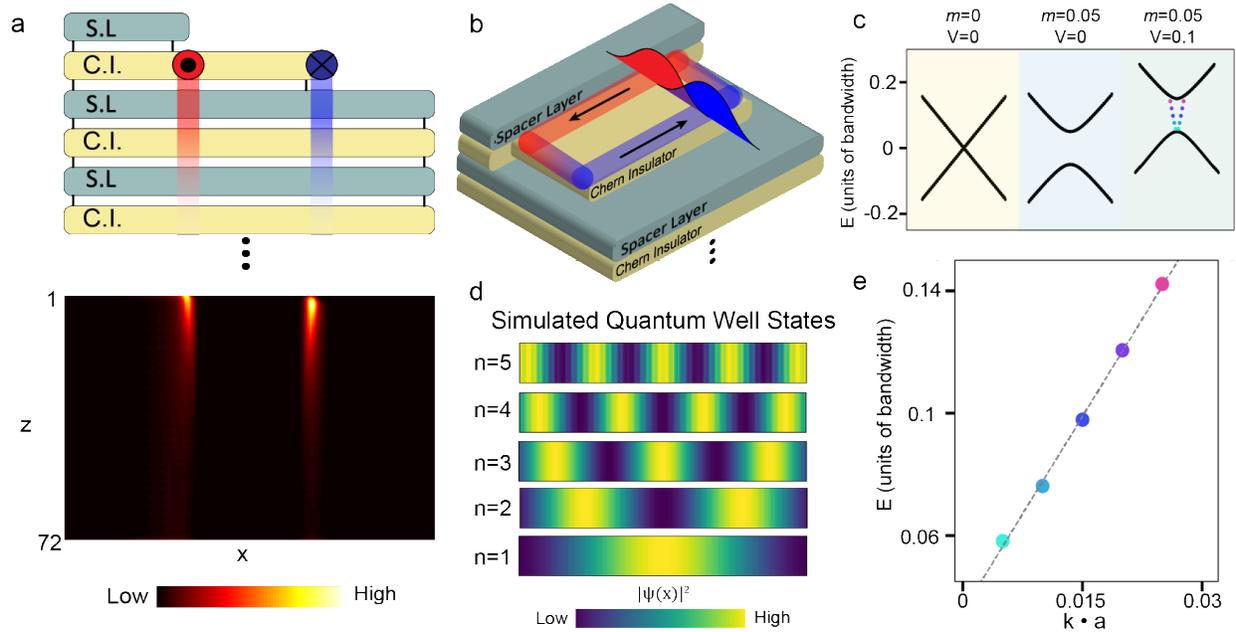

**a**, Schematic of coupled Chern insulator layers with alternating spacer layers, with a Chern insulator terrace (top panel), a simple stack that mimics our geometry. The red and blue circles indicate 1D chiral modes travelling out of and into the page respectively, with the red and blue trails indicating their power law decay into the bulk. The bottom panel is a numerical simulation of 70 full Chern insulator layers plus two partial layers as shown the top panel. The bright regions in the bottom panel indicate chiral edge states on the exposed terrace as a result of the numerical simulation. **b,** Schematic of the narrow $Co_3Sn$ terraces observed applied to this model. The thinly exposed Chern insulator layer ($Co_3Sn$ plane) will host chiral edge states, and when the terrace is sufficiently narrow, the two counterpropagating states will overlap and interact, as shown by the red and blue curves. **c,** Low momentum dispersion for different parameters of *m* and *V*. With no mixing or potential well, two linearly dispersing bands cross at zero momentum (light yellow left panel). Adding a mixing term of five percent of the bandwidth, a gap opens as the two bands hybridize (light blue middle panel). Further adding a potential of ten percent of the bandwidth increases the energy of all states and causes five states to be confined entirely within the potential well (light green right panel), indicated by the colored dots within the gap. **d,** Wavefunctions of the confined, linearly dispersing states shown in **e**. **e,** Dispersion of the states confined within the potential well from the light green panel in **c**. A straight dashed line is provided as a guide to the eye.

Supplemental Information

# Supplemental Information 1 | Model of Chern Insulator Bilayer and Weyl Semimetal

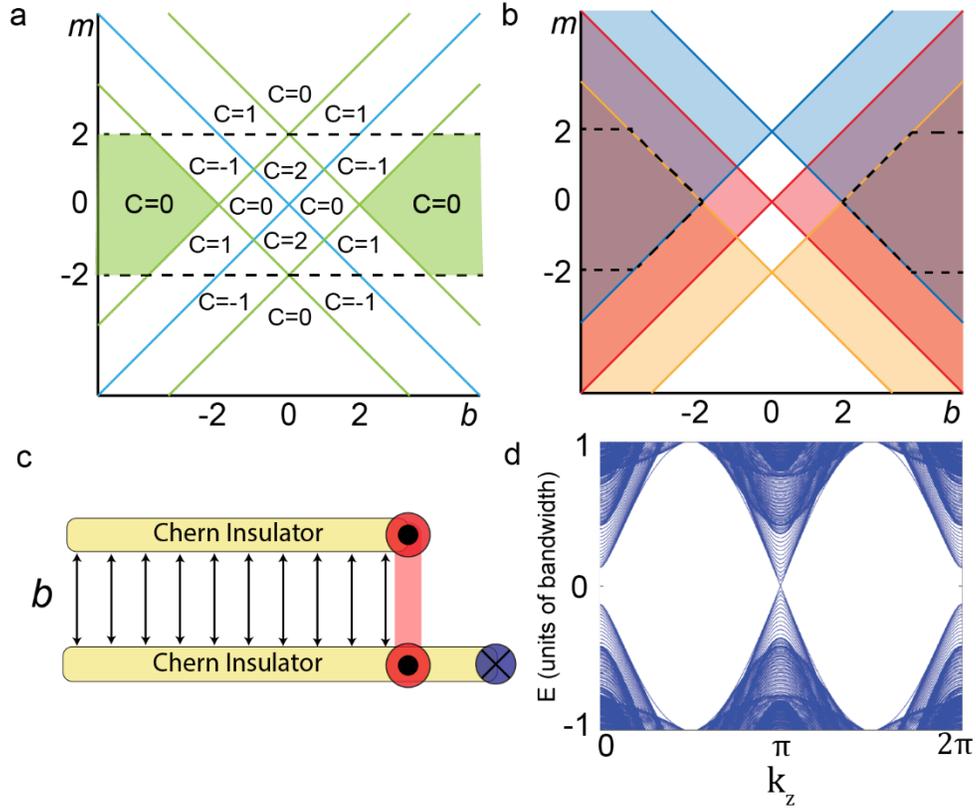

**a,** Phase diagram for the Chern bilayer system. In the following $b$ represents the coupling between two layers and $m$ is the mass parameter. The horizontal dashed lines represent the region in which a single layer is a Chern insulator when $b = 0$ and $m$ is varied. Diagonal lines represent phase boundaries ($m = \pm b$ is blue, $m = -2 \pm b$ and $m = 2 \pm b$ are green) where the Chern number changes. The green shaded regions indicate coupling where each layer in the bilayer is trivial, but a single layer would be topological without coupling. **b,** Weyl semimetal phase diagram for a stacking of Chern insulators with coupling. Each of the shaded regions (blue, red, and yellow) represent a condition for Weyl nodes ($-1 < \frac{-2-m}{b} < 1$, $-1 < \frac{-m}{b} < 1, -1 < \frac{2-m}{b} < 1$) being satisfied. **c,** Schematic of the Chern insulator bilayer with a step-like geometry as viewed from the side, showing localized chiral edge modes at the steps. The mode present at the beginning of the bilayer exists on both surfaces due to the coupling. **d,** Dispersion of bands within a 8-layer coupled Chern insulator system (m=0.75, b=5) similar to 72 layer system shown in Fig 4a. The low energy chiral edge modes localized at steps can be seen at E=0, $k_z = \pi$.

**Bilayer system:** We will begin by considering a model having two Chern insulator layers. The Bloch Hamiltonian we will consider is

$$H(k) = \sin k_x\, I \otimes \sigma^x + \sin k_y\, I \otimes \sigma^y + (m + \cos k_x + \cos k_y) I \otimes \sigma^z + b \tau^x \otimes \sigma^z$$

where $\tau^a$ are Pauli matrices in the layer space, and $\sigma^a$ represent spin. This Hamiltonian has four energy bands with energies

$$\pm E_\pm(k) = \pm\sqrt{\sin^2 k_x + \sin^2 k_y + (M(k) \pm b)^2}$$

where $M(k) = m + \cos k_x + \cos k_y$. The parameter $b$ represents the coupling between two layers, and we see that it effectively shifts the mass parameter $m$. For $b=0$, the system has critical points at $m$= -2, 0, 2 where the Chern number changes its value. For finite $b$ there are critical lines in the $(b,m)$ plane given by $m = -2 \pm b$, $m = \pm b$, $m = 2 \pm b$ as represented in **a**.

Interestingly, when examining the phase diagram, we find regions where the coupling can drive a pair of non-trivial Chern insulators into a completely trivial phase (green shaded region in **a**). This region of the phase diagram has the remarkable feature that a complete bilayer will not exhibit any topological edge states, but a step-like geometry can exhibit a pair of counter-propagating chiral modes. Thus, a trivial stepped bilayer can exhibit a plateau with edge states in a wide swath of parameter space. The idea is that by themselves each layer is topological, but when coupled they are trivial. Consequently, when a part of a layer is removed, the exposed layer below is now "topological" again. The region in **a** that is shaded green is exactly the region where this occurs because the total Chern number is zero and both layers are trivial, but $m$ is in the regime where, if $b$ was off, the system would be topological. The only thing remaining is to see if this survives when coupled to many layers that act to form a Weyl semimetal.

**3D Weyl semimetal:** We can make a modification to our bilayer Bloch Hamiltonian to represent a 3D Weyl semimetal. Essentially, the tunneling term between layers needs to be extended to include many layers stacked in the z-direction, and we obtain

$$H(\mathbf{k}) = \sin k_x\, \sigma^x + \sin k_y\, \sigma^y + (m + \cos k_x + \cos k_y + b \cos k_z)\sigma^z$$

where again $\sigma^a$ represents spin and $b$ is the coupling amplitude in the z-direction. If $b$ is weak, and $m$ is tuned so that the layers are nominally in a non-trivial Chern insulator phase, then the system will form a 3D weak topological insulator with sheets of chiral surface states on the xz and yz surface planes. If $b$ is strong enough to close the 2D bulk gap, then the system will form a Weyl semimetal phase with two Weyl nodes separated in the $k_z$ direction. The regions of the phase diagram in the $(b, m)$ plane that represent Weyl semimetal phases are those for which a non-trivial solution of $m + b \cos k_z = -2, 0, 2$ can be found. We thus have the conditions

$$-1 < \frac{-2 - m}{b} < 1$$

$$\text{or } -1 < \frac{-m}{b} < 1$$

$$\text{or } -1 < \frac{2 - m}{b} < 1.$$

When at least one of these three conditions are met, the system will have Weyl nodes. If the first (second) [third] condition is met, there will be Weyl nodes at $(k_x, k_y, k_z) =$ $(0,0, \pm k_z^c), ((\pi, 0, \pm k_z^c), (0, \pi, \pm k_z^c)), [(\pi, \pi, \pm k_z^c)]$ for some value of $k_z^c$. These are illustrated by the shaded regions in the $(b, m)$ phase diagram in **b**.

Importantly, we see that the region in which we expect a Chern insulator bilayer to exhibit localized modes on step geometries is completely within a Weyl semimetal phase. If we explicitly solve a Weyl semimetal Hamiltonian (m=0.75, b=5) with this step geometry using numerical exact diagonalization, we find a spectrum shown in **d** (calculated for 8 layers) and localized chiral modes on the step plateau as shown in Fig 4a (calculated for 72 layers). Note that these results are qualitatively insensitive to the number of layers used in the numerical model. The modes are localized at the steps but penetrate the layered bulk since the interior is nominally a gapless Weyl semimetal.

# Supplemental Information 2 | Determination of Surface Composition from Layer Below Defect Symmetry

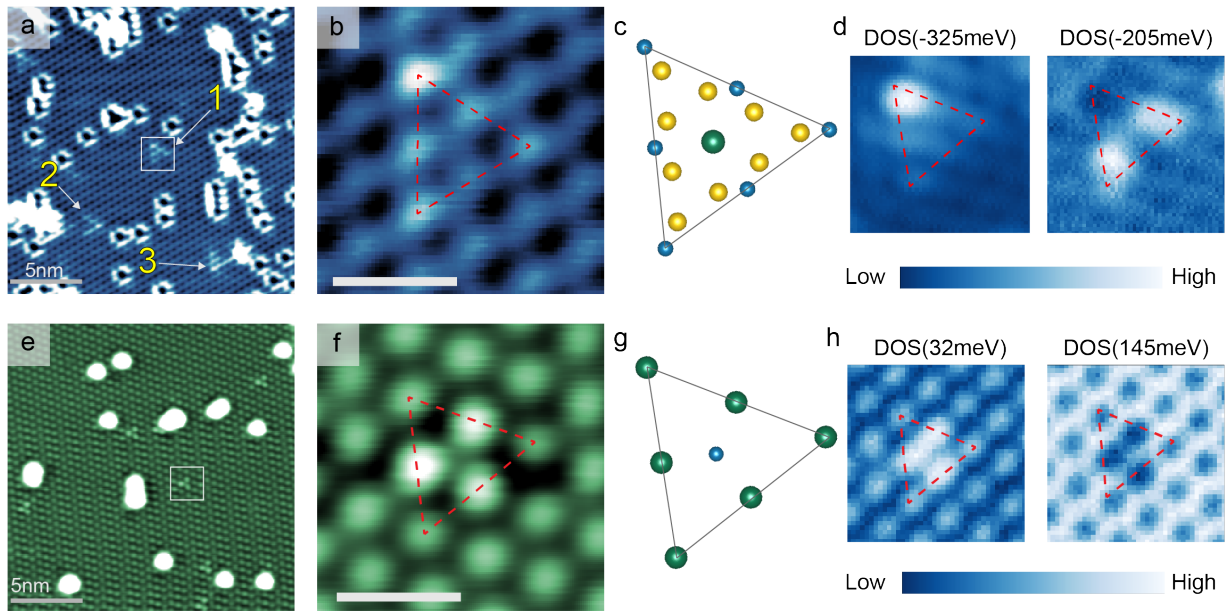

**a,** 20 nm x 20nm topography of the surface identified from Fig 1g showing directional triangular DLBs. Three distinct defects with the bright vertex pointing up, right, and left are indicated by numbers 1, 2, and 3 respectively. **b,** Region within white box in **a** showing a zoomed in triangular DLB. The red dashed line is provided for comparison with **c**. The inset scale bar is 1nm. **c,** Schematic of the triangle in **b** as the top layer S (blue spheres) and the $Co_3Sn$ layer below. **d,** DOS maps of region in **b** at -325 meV and -205 meV showing that the density of state signature of the triangular DLB has a reduced symmetry compared to two shifted hexagonal layers, indicative of the $Co_3Sn$ layer below. **e,** 20 nm x 20 nm topography of the surface identified from Fig 1h showing clover DLBs. All DLBs found on this surface were identical. **f,** Region within white box in **e** showing a zoomed in clover DLB. The red dashed line is provided for comparison with **g**. The inset scale bar is 1nm. **g,** Schematic of the triangle in **g** as the top layer Sn (green spheres) and the S layer below. **h**, DOS maps of region in **f** at 32 meV and 145 meV showing a trifold symmetry that is consistent with a Sn surface and a S layer below.

For the $Co_3Sn_2S_2$ samples, since cleavage most often occurs between the Sn layer and S layers, the exposed surface imaged with STM will either be the hexagonal S layer with the kagome $Co_3Sn$ layer directly beneath, or hexagonal Sn with hexagonal S directly beneath. As the kagome layer has a different structure compared to the hexagonal S layer, the symmetry of the density of states signatures from defects in the layer below (DLB) can be used to distinguish the Sn surface from the S surface.

DLBs can be identified in the topographies (Supp. Fig. 2a and Supp. Fig 2e) as extended defects centered in between the top layer atoms. On surfaces with vacancies, we observe large triangular DLB features occupying three lattice sites on each side, showing one vertex that is clearly brighter than the other two (Supp. Fig. 2b). As seen in the topography, the position of the bright vertex can be different for different defects (labelled 1,2,3 in Supp. Fig. 2a). The distinct bright vertex can also be seen in the real space density of state maps, obtained in the vicinity of these defects (Supp. Fig. 2d). Considering the layers beneath the

S or Sn planes, we realize that only vacancies or substitutional impurities at the Co sites of the $Co_3Sn$ kagome plane have the right symmetry to give rise to the observed density of states signature. This explains both why the whole triangular DLB appears in one angular orientation (i.e., not rotated 60 degrees), and why there are three different bright vertices. On the surface with the adatoms, we do not observe these triangular DLBs. Instead, a "clover" DLB is seen, composed of three equally bright adjacent atoms as shown in Supp. Fig. 2e,f which persists in DOS maps from -250mV to 250mV (Supp. Fig. 2g,h). The symmetry of these clover DLBs is consistent with a Sn surface showing defects in the S sites in the layer below. With these two pieces of evidence, we unambiguously identify the surface containing vacancies (Fig. 1g) to be the S surface and the surface containing adatoms (Fig. 1h) to be the Sn surface. This determination is of specific importance for future studies of this material, as the nature of Fermi arc states is dependent on the local surface potential.

## Supplemental Figure 3 | Point Spectra on S and Sn Step Edges

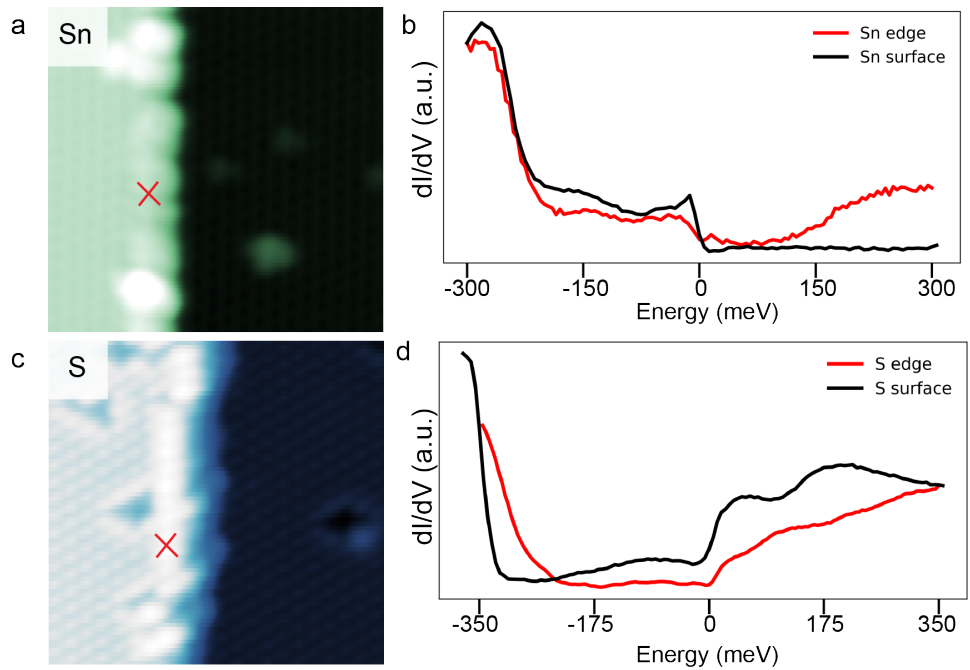

**a,** Topography (10nm x 10nm) of a Sn step edge. **b,** Comparison of the spectra on the Sn surface to a spectrum on the Sn edge. The edge spectra was taken at the red cross shown in **a**. **c,** Topography (10nm x 10nm) of a S step edge. **d,** Comparison of the spectra on the S surface to a spectrum on the S edge. The edge spectrum was taken at the red cross shown in **c**.

Supplemental Figure 4 | Additional Analysis of Co$_3$Sn Edge State

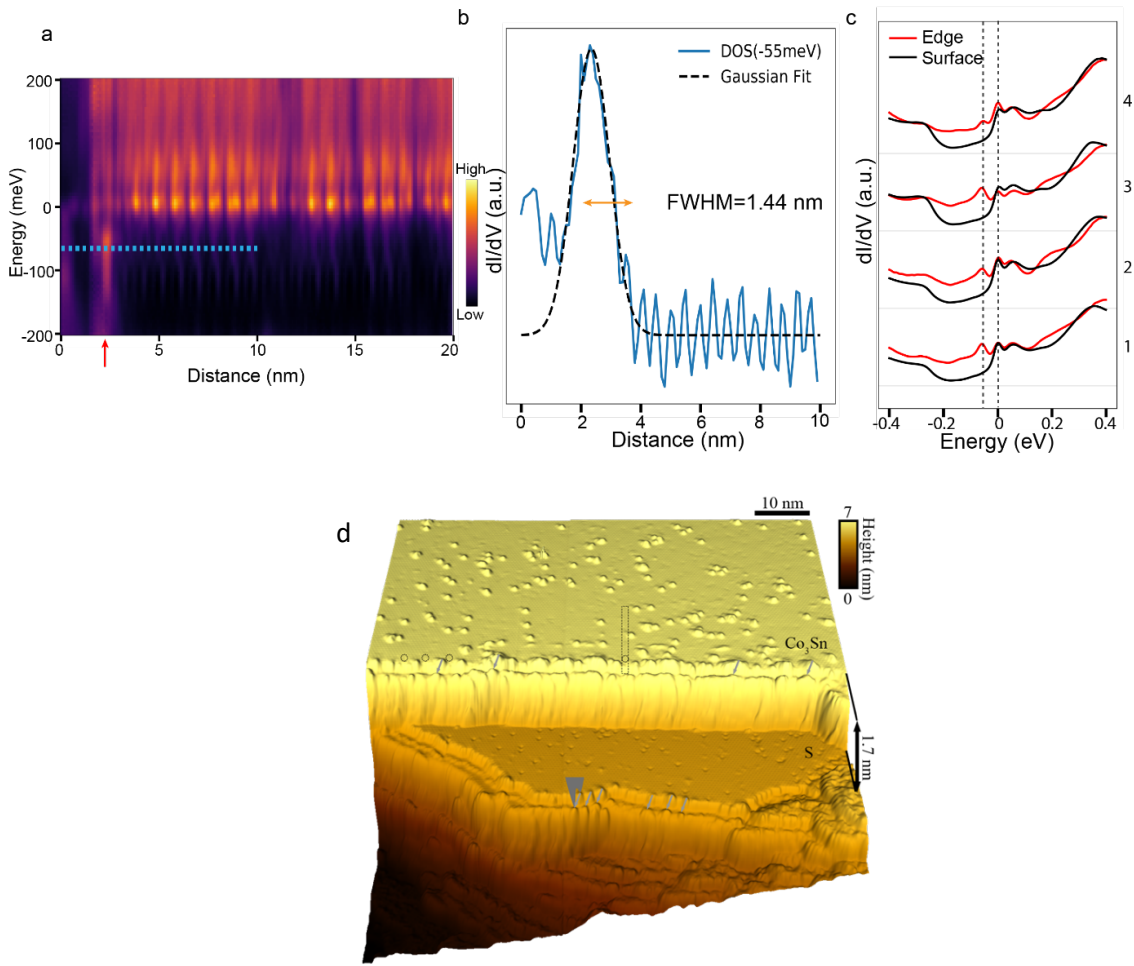

**a,** Spectroscopic heatmap from Fig. 2c in the main text, with a blue dashed line to show the spatial evolution of the DOS at -55meV in **b**. **b,** Spatial dependence of the DOS at -55meV across Co$_3$Sn edge. The peak in the DOS at -55meV is from the edge state. The peak is fit with a gaussian line shape, and a full width half maximum (FWHM) of 1.44nm is found. This gives us an estimate for the spatial extent of the edge state. **c,** Comparison of four Co$_3$Sn edge spectra and nearby surface spectra. The spectra labelled 1 are the pair shown in Figure 2 of the main text. All edge spectra show an enhancement of the density of the states below the Fermi energy, peaking at approximately -60 meV, when compared to spectra taken away from the edge. Vertical dashed lines at 0 meV and -60 meV are shown for comparing spectra. **d,** A larger scale image of the Co$_3$Sn terrace and step where the image in Fig.2a in main text was obtained (dotted area). There are many defects on the edge where these spectra are taken, but no confinement effects are seen in these spectra. This is in contrast to the terrace quantum well like states seen in Figure 3 and Supplemental Information 5, where the close proximity of another edge state allows for hybridization and confinement.

# Supplemental Information 5 | Two Other Co$_3$Sn Terraces with Linearly Dispersing Quantum Well Like Bound States

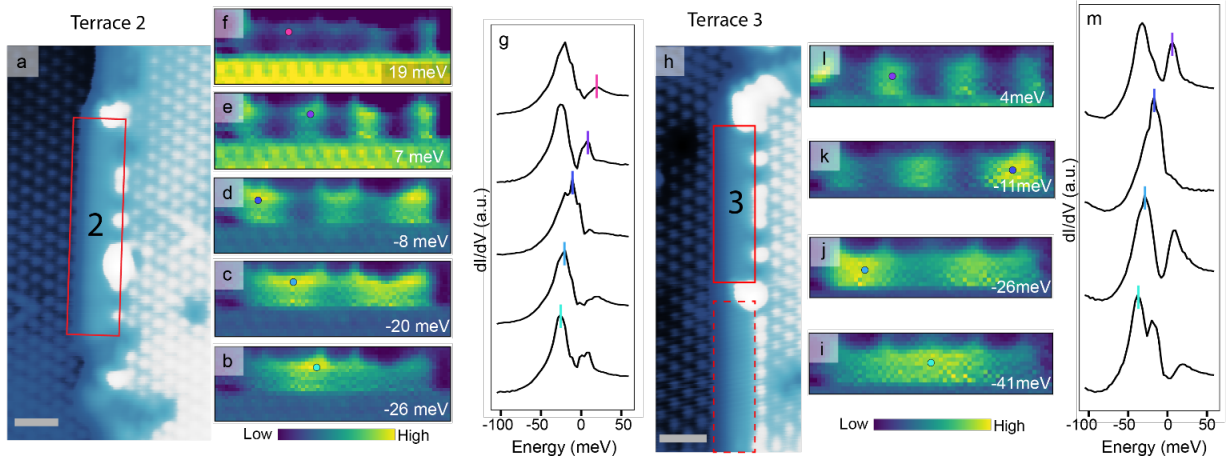

**a,** Topography of second terrace region showing quantum well like behavior. Scale bar in bottom left corner is 2 nm. Terrace 2 is approximately 6.1 nm long. Two triangular DLBs can be seen both on the top and bottom surface. In general, the terrace length is determined by the spatial extent over which we see the `quantum well like' states. **b-f,** DOS images of n=1 to n=5 quantum well like states for terrace 2. Region shown is depicted by red rectangle containing 2 in **a**. Location of spectra shown in **g** are indicated by small colored circles. The color indicates the n$^{th}$ bound state with same scheme as in Fig 3. Note that for n=1 we used the distance to the left edge to calculate the wavelength due to the presence of an impurity on the right. **g,** Spectra at local density of states maxima indicated in **b-f**. Spectra are offset for clarity. States are represented as circles in Fig 3h. **h,** Topography of two terrace regions showing quantum well like behavior. The red dashed box is same terrace shown in Fig 3a, while the red box containing 3 is the 5 nm long region containing quantum well like states in **i-l**. Scale bar in bottom left corner is 2nm. **i-l,** DOS images from n=1 to n=4 in terrace 3. Region shown is depicted by red rectangle containing 3 in **h**. Location of spectra in **m** indicated by small colored circles. **m,** Spectra at local density of states maxima indicated in **i-l**. Spectra are offset for clarity. States are represented by triangles in Fig 3h.

# Supplemental Information 6 | Single Row of Atoms on Terrace and Comparison to Other Surfaces

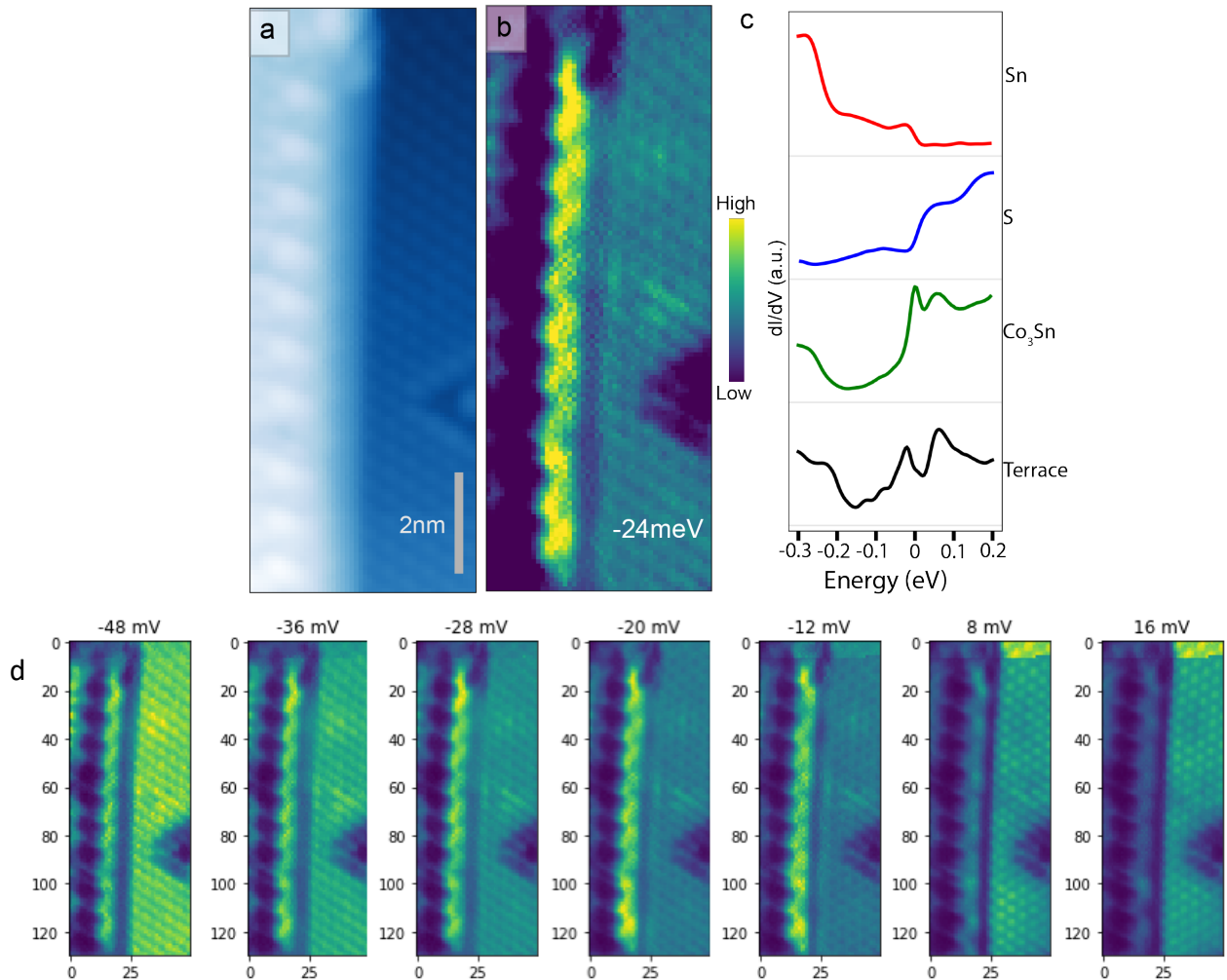

**a,** 5nm x 13nm topography of a Co₃Sn terrace at a S step edge with only one row of atoms in Co₃Sn layer exposed. Gray 2nm scale bare in the lower right. **b,** DOS image at -24mV. The contrast at this bias value allows for better identification of the exposed Co₃Sn layer. **c,** Typical spectra seen on the Sn surface (red), S surface (blue), Co₃Sn surface (green), and the terrace/bright region shown in **b** (black). The spectra are offset for clarity, with a gray line representing zero differential conductance for each spectrum. **d,** DOS image at a series of voltages showing the distinct lack of quantum well like states.

Here we report a single row of exposed atoms below a S surface. The single row of atoms is visible as the bright region in the density of states map shown in **b**. While a single row of atoms is not enough to topographically identify this plane, the spectra observed on this terrace (as shown in **c**) most resembles that found on the Co₃Sn plane (also see Fig.2). Unlike terraces observed with larger widths, no quantum well like bound states were observed on this terrace. This is consistent with description of two counter-propagating edge states, since they would annihilate each other if directly on top of one another.

# Supplemental Figure 7 | Analysis of Terrace Height Difference for Layer Determination

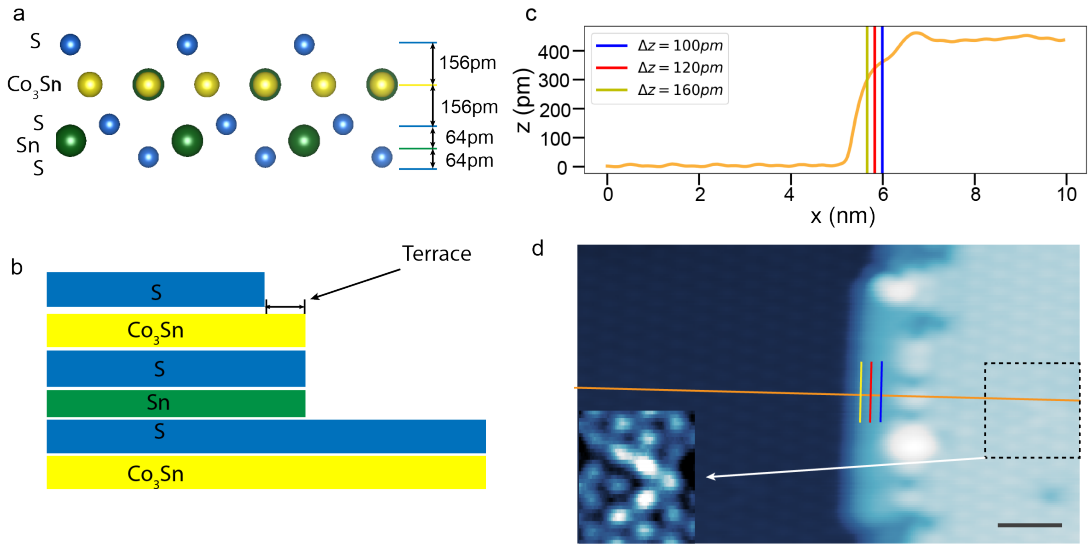

**a,** Height differences between S, $Co_3Sn$, and Sn layers. **b,** Schematic of the determined terrace surface composition, with a small thin terrace of $Co_3Sn$ exposed. **c,** Height determination of the terrace from a z line profile. Due to the narrow width of the terrace, the measured height changes across the terrace. This is indicated by three colored lines (blue, red, and yellow), which give height changes of 100pm, 120pm, and 160pm respectively. **d**, Larger area topography of the terrace where the quantum well like bound states in Terrace 2 were observed. The top layer can be identified from the defects and the spectra as the S plane. The scale bar is 2nm. The orange line is the location of the line profile shown in **c**. The blue, red, and yellow lines are the locations along the line profile used in height determination in **c**. The dashed box on the right is the location of a triangular layer below defect shown in the inset. The existence of the triangular layer below defect confirms that the top layer is the S layer with $Co_3Sn$ directly beneath.

Here we perform a detailed analysis of the height change from the S surface to the terrace and find that the height difference observed is approximately 130pm ± 30pm. The distances between the S and $Co_3Sn$ layer directly below is 156pm. Crucially the distance to the next layer, a S layer, is 312pm which too large to account for the change in height observed. The experimental error in our height determination comes from the thin terrace width as well as density of states contributions.

Supplemental Figure 8 | Discrete Model of Linearly Dispersing Modes in Finite Quantum Well

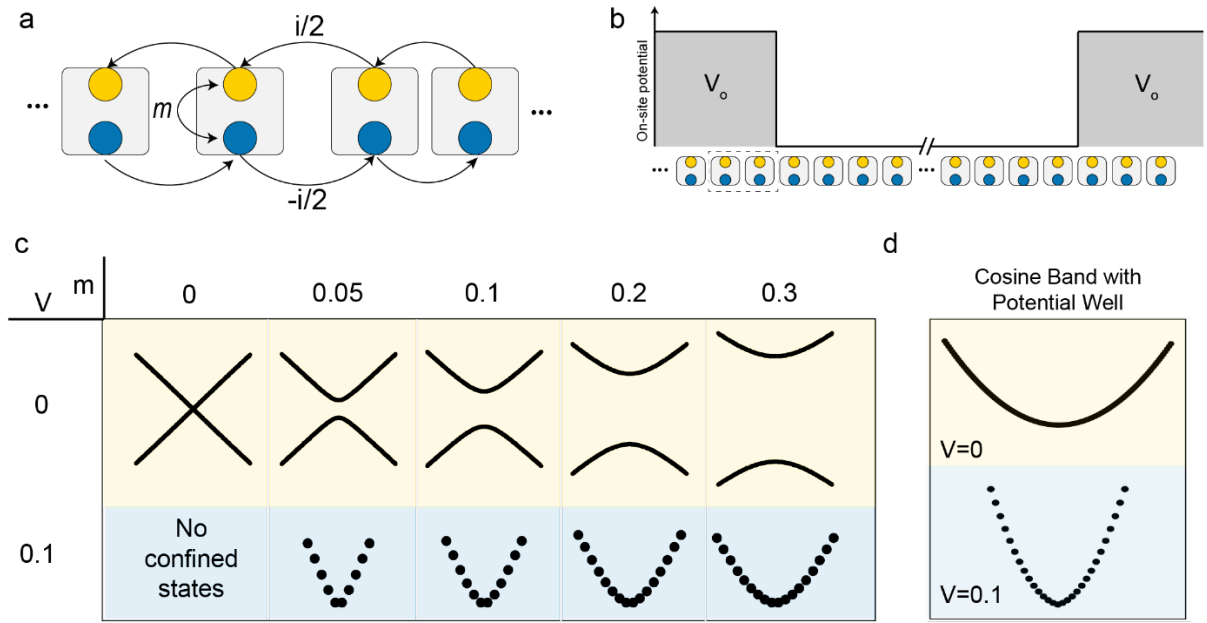

**a**, Schematic of two unit cells within tight binding Hamiltonian modeling this system: $H = \sum_n \frac{i}{2} c^\dagger_{n+1,\alpha} c_{n,\beta} \sigma^z_{\alpha\beta} - \frac{i}{2} c^\dagger_{n,\alpha} c_{n+1,\beta} \sigma^z_{\alpha\beta} + m c^\dagger_{n,\alpha} c_{n,\beta} \sigma^x_{\alpha\beta} + V(n) c^\dagger_{n,\alpha} c_{n,\beta} \delta_{\alpha\beta}$. This is similar to Fig. 4c, but is reproduced here for context with **b**. A rectangle indicates one unit cell containing two cites represented by yellow and blue circles. Hopping can occur between the same site on adjacent unit cells. The phase relation is such that hopping on one site has positive linear dispersion around zero momentum (yellow circles) while the other site has a negative linear dispersion around zero momentum (blue circles). The sites in one unit cell are allowed to interact via a mixing term *m*. **b**, Schematic of the potential well $V(n)$. An on-site potential much smaller than the bandwidth is applied to states outside the well. For **c** and **d**, a well size of 100 out of 1000 sites is used. **c,** Low momentum band dispersion for the model in **a** for various mixing strengths (light yellow panels) and the states confined within a potential well of 0.1 (light blue panels). For no mixing, no linearly dispersing states are confined within the potential well. As the mixing increases, the low energy and bands become more quadratic and the confined states mimic this behavior. The energy of all states would increase with a potential, as seen in Fig. 4e, although the states outside the well are not shown here. The blue panels are scaled and shifted relative to the yellow panels for easier comparison. **d**, Low momentum band dispersion for a single cosine like band (light yellow panel) and the states confined within a potential well of 0.1 (light blue panel) for comparison with our model.